# Measurement of discrete energy-level spectra in individual chemically-synthesized gold nanoparticles

*Ferdinand Kuemmeth, Kirill I. Bolotin, Su-Fei Shi, Daniel C. Ralph*[*]

Laboratory of Atomic and Solid State Physics, Cornell University,

Ithaca, New York 14853

*ralph@ccmr.cornell.edu

We form single-electron transistors from individual chemically-synthesized gold nanoparticles, 5-15 nm in diameter, with monolayers of organic molecules serving as tunnel barriers. These devices allow us to measure the discrete electronic energy levels of individual gold nanoparticles that are, by virtue of chemical synthesis, well-defined in their composition, size and shape. We show that the nanoparticles are non-magnetic and have spectra in good accord with random-matrix-theory predictions taking into account strong spin-orbit coupling.

Nanometer-scale metal particles are small enough that their discrete spectra of quantum-mechanical electron-in-a-box energy levels can be resolved using electron tunneling spectroscopy at low temperature. Several different methods have been developed for making electrical contact to single nanoparticles and measuring their spectra, including devices made with nanoconstrictions,[1,2] shadow evaporation,[3] and electromigration,[4] as well as scanning-tunneling-microscope studies.[5] Detailed analysis of the electronic spectra in metal particles has shown that they are not well described by simple models of free, non-interacting electrons, but rather they are affected by all of the different types of forces and interactions that govern the electronic structure within metals. This has allowed the spectra to be used for making detailed studies of superconducting pairing, spin-orbit coupling, ferromagnetic exchange interactions, and other spin-dependent effects in metals.[6] The metal nanoparticles studied within previous fabricated devices were generally formed by the clustering of metal atoms evaporated onto an insulating surface, with oxide tunnel barriers on top of the particles. These experiments therefore lacked good control over the particles' size and shape, their tunnel resistances were highly variable, and they were often affected by background charge fluctuations that limited the quality of the spectral measurements.

Here we demonstrate the use of "bottom-up" fabrication techniques to assemble individual chemically-formed gold nanoparticles within single-electron transistors, with a monolayer of organic linker molecules forming the tunnel barriers between the nanoparticle and the transistor electrodes. We show that the devices are sufficiently stable to make detailed measurements of the electron-in-a-box energy levels. In fact, the spectra are of extremely high quality -- we resolve more levels in the excited-state electronic spectrum for a fixed gate voltage (as many as 40), with less background noise, than in any other measurement on a quantum-dot system as far as we are aware. The large number of levels resolved indicates that our chemically-formed nanoparticles may have fewer background charge fluctuations than in previous metal-nanoparticle experiments and also suggests[7] that the chemically-formed nanoparticles may contain less static disorder. We are able to study the changes in the electron spectrum as electrons are added one by one by tuning a gate voltage, and we can make comparisons to random matrix theory (RMT) predictions for the level statistics of the excited states and their magnetic-field responses. We find good agreement with RMT predictions for the regime of strong spin-orbit coupling and ballistic transport. In previous work, chemically-formed semiconductor and metal nanoparticles have been incorporated into similar devices to study Coulomb blockade physics,[8,9,10,11,12] but such devices have not been used for measurements of the electron-in-a-box states in metals.

We study gold nanoparticles of approximately spherical shape (Fig. 1a, inset) synthesized from an aqueous solution of hydrogen tetrachloroaurate ($HAuCl_4$) at boil by chemical reduction using the recipe of ref. 13. We control the diameter of the particles (5-15 nm) by the amount of reducing agent added (a mixture of tannic acid and trisodium citrate). The resulting ruby-red colloidal solution contains approximately $10^{11}$ negatively charged nanoparticles per ml and shows size-dependent absorption peaks in the ultraviolet (~260 nm) and in the green (~530 nm) (not shown). Structural characterization using electron diffraction imaging in a transmission electron microscope (Fig. 1b) reveals that the nanoparticles contain grain boundaries but are otherwise crystalline. Small deviations from spherical symmetry ($\geq$ the Fermi wavelength $\sim$ 0.5 nm) remove orbital degeneracies and are expected to eliminate any possible shell structure in the nanoparticles' electronic spectra. The colloidal suspensions are stored in the dark and are stable over weeks against precipitation due to electrostatic repulsion between the negatively charged nanoparticles.

The single-electron transistors (SETs) that we use for performing electron tunneling spectroscopy of the nanoparticle electronic levels are prepared similarly to SETs fabricated with evaporated nanoparticles,[4] but instead of an aluminum gate we use a degenerately-doped silicon backgate (with 30 nm of thermal oxide) that is not harmed by the low pH present during self-assembly. On top of the silicon backgate, we use electron-beam lithography and lift-off to pattern bow-tie shaped gold wires (16 nm thick and approximately 100 nm minimum width). After cleaning the chips using acetone, isopropanol and an oxygen plasma, the wires are broken using electromigration in a room temperature probe station to form gaps a few nm wide separating source and drain electrodes.[14,15] Each chip is submerged in 0.05% (w/v) aminoethylamino-propyl-trimethoxysilane (APTS, Sigma#66668, $(CH_3O)_3Si(CH_2)_3NH(CH_2)_2NH_2$) for 10 minutes to form a self-assembled monolayer, then rinsed in DI water. The silane endgroup of APTS should bond strongly to the $SiO_2$ surface of our substrate; in addition we believe that the APTS attaches to the gold electrode as well because ultimately we observe similar surface densities of nanoparticles assembled onto both the $SiO_2$ and Au. After baking the chip at 120ºC for 30 minutes, the resistance of each break junction is recorded (typically > 10 GΩ). The chip is then submerged in the gold colloid solution for at least 12 hours. We control the density at which the nanoparticles self-assemble onto the surface by adjusting the solution pH to partially protonate the amino groups of the APTS, thereby attracting the negatively-charged gold nanoparticles to the surface (Fig. 1c).[16,17] The density of nanoparticles shown in Figure 1d was achieved at pH ≈ 2 by addition of citric acid (1 g/200 ml).[18] The chip is then rinsed in DI water and dried. Typically 30% of the junctions display a drop in room-temperature resistance to 0.1-5 GΩ after the nanoparticle deposition step, and half of these display well-defined Coulomb-blockade characteristics at 4.2 K. These devices are selected for further study at dilution-refrigerator temperatures, where the discrete electronic energy-level spectrum can be resolved (Fig. 1a). All of the data we present were taken at dilution refrigerator temperatures (electron temperature ≤ 90 mK).[19]

If the density of the nanoparticles in a device is not too large, electrical current flows through an individual nanoparticle only. This is determined by the presence of regularly-shaped regions of zero conductance (Coulomb blockade diamonds)[20] in measurements of the differential conductance $dI/dV_{SD}$ over large ranges of bias and gate voltage (see Fig. 2a for device #3) and can be confirmed using inspection by scanning electron microscopy after transport measurements are completed (Fig. 1d). The size of the Coulomb blockade diamonds and the slope of the thresholds for $dI/dV_{SD}$ allow a determination of the capacitances $C_S$, $C_D$, and $C_G$ of the nanoparticle to the source, drain, and gate. For device #3 in Fig. 2, we find $C_S$ = 1.1 aF, $C_D$ = 1.1 aF, and $C_G$ = 0.03 aF, so that the charging energy is $E_C = e^2/2(C_S+C_D+C_G)$ = 36 meV.

We perform detailed spectroscopic measurements of the electron-in-a-box quantum states of each nanoparticle by sweeping the bias voltage $V_{SD}$ while recording the direct current $I$, and then stepping the values of the gate voltage $V_G$ or magnetic field $B$. When plotting the differential conductance $dI/dV_{SD}$ vs $V_G$ and $V_{SD}$ (Fig. 1a, Fig. 2b) the discrete energy levels of the nanoparticle appear as parallel bright lines near the "degeneracy points" -- the values of $V_G$ at which adjacent charge states have the same energy, so that current flow is possible for small values of bias (at $V_G$ = -2.3, 3.1, and 8.5 V in Fig. 2a). Lines with positive and negative slope correspond to tunneling transitions across the two different tunnel barriers to the source and drain electrodes. If $N$ and $N$+1 are the numbers of electrons in the two near-degenerate charge states between which tunneling transitions occur at low bias, for a given sign of bias the lines with different slopes correspond to $N \rightarrow N+1$ and $N+1 \rightarrow N$ transitions. In contrast to previous experiments on nanoparticles with oxide tunnel barriers, in the majority of our devices the spectrum of excited-state tunneling transitions associated with the source electrode and the spectrum associated with the drain are approximately equally prominent, indicating that that the two molecular tunnel barriers have resistances that are the same to within roughly a factor of three. (Device #2 discussed below is an exception.)

For the spectrum in Fig. 2b, the mean level spacing for a given charge state at $B = 0$ is (after converting from source-drain voltage to energy by taking into account the capacitive division of voltage across the two tunnel junctions[1]) $\langle \delta \rangle = 0.33$ meV and the standard deviation in the level spacing is $\Delta \delta = 0.10$ meV. The value of $\langle \delta \rangle$ is consistent with expectations for the mean diameter of the nanoparticles as measured by TEM prior to device fabrication; for this batch of nanoparticles the diameters were 9.1 nm ± 10%, implying a mean level spacing of 0.32 meV. Typically, the mean level spacings measured for different nanoparticles from the same synthesis vary within a 25% range. Figure 1a shows that more than 40 separate resonances in the differential conductance can sometimes be resolved. This is more excited-state levels than have been resolved in any other semiconducting or metallic quantum dot system, as far as we are aware.

In the remainder of this Letter we explore the nature of these spectra and show that (*i*) the tunneling rates display random state-to-state fluctuations as expected for the wavefunctions of a chaotic quantum dot, (*ii*) the discrete spectrum can be described as arising from the filling of single-particle levels with no discernible "scrambling" by variations in electron-electron interactions as electrons are added one by one, and (*iii*) each quantum state is affected differently by strong spin-orbit coupling, in good agreement with RMT predictions for ballistic gold particles.

(*i*) The electron wavefunctions in the gold nanoparticles near the Fermi level should be highly oscillatory with large fluctuations in magnitude, because the Fermi wavelength in gold is approximately 0.5 nm, much smaller than the diameter of the nanoparticles. These oscillations are expected to lead to fluctuations in tunneling matrix elements, thereby producing variations in tunneling rates when comparing different energy levels, and also differences in the tunneling rates to the source and drain electrodes for a given energy level, despite the fact that the average tunneling resistances between the nanoparticle and the two electrodes can be approximately equal.[21,22] We find that there are, indeed, large fluctuations in tunneling rates between different energy levels, as is evident in the large variations in the magnitudes of the differential conductance for the different tunneling resonances (Fig. 2c). Differences in the tunneling rates for a given energy level to the source and drain electrodes can also be seen. As expected for a device with symmetric tunnel barriers, each energy level that gives a resonance at positive $V_{SD}$ also gives a resonance at negative bias, so that line cuts obtained along positive and negative $V_{SD}$ agree with each other in terms of peak positions (energy). However, the line cuts can be strikingly different in terms of conductance peak heights (related to tunneling rates) (Fig. 2c). For example, the quantum state giving rise to the resonance marked by a solid black circle has a larger tunnel probability to the source than to the drain, whereas the situation is reversed for the next higher quantum state (open black circle). The resulting asymmetry in current can be understood as follows (Fig. 2c, insets): at high bias and for $V_G$ tuned well below the degeneracy point, the rate-limiting step for the tunnel current in Fig. 2 is always tunneling onto the particle (as tunneling off the particle can occur via many channels), and for a quantum state which is spatially quasi-random this limiting step can occur with drastically different rates depending on whether the electron tunnels from the source or the drain electrode.

(*ii*) Next we show that the electrons in our gold nanoparticles can be described, to a good approximation, as effectively non-interacting quasiparticles which fill doubly-degenerate (at $B = 0$ T) single-particle levels whose relative energies do not depend on the charge state of the nanoparticle. In other words, variations in electron-electron interactions are sufficiently weak that the interactions can be accounted for entirely by state-independent charging energies; interactions do not cause the underlying electronic spectrum to be scrambled significantly as electrons are added one by one to the nanoparticle using a gate voltage.[22,23] Figure 3a compares the differential conductance spectrum obtained near one degeneracy point of device #1 (tunneling transitions between electron numbers $N$ and $N+1$, grayscale) with that obtained from an adjacent degeneracy point in the same device (tunneling transitions between electron numbers $N+1$ and $N+2$, pink color scale). Except for the lowest-energy state which is available for tunneling in the pink spectrum but which is unoccupied by an electron and therefore does not give a tunneling signal in the grayscale spectrum, the two spectra match. This is true even though a charge $-e$ has been added to the nanoparticle and the gate voltage has been changed by more than 10 Volts. We believe that this insensitivity arises because the hardwall confining potential seen by the electrons in our nanoparticles causes the volume of the quantum dot to be independent of the charge number (unlike 2-dimensional semiconducting quantum dots defined by metal gates[23] but similar to semiconductor dots defined by local oxidation[24]), because the screening length is much smaller than the gold nanoparticle diameter, and because exchange interactions in the noble metals are predicted to be very weak.[25,26] Energy shifts and splittings due to state-dependent electron-electron interactions have been observed previously in nanoparticles with much smaller diameters, and in nanoparticles made of metals with stronger interactions.[3,19,27]

In Fig. 3b we show the magnetic-field dependence of the tunneling resonances for the $N+1 \rightarrow N$ transitions (grayscale) and the $N+2 \rightarrow N+1$ transitions (pink color scale) in device #1, corresponding to tunneling of an electron from occupied states on the nanoparticle to the drain electrode. Again, except for the first (lowest $|V_{SD}|$) transition in the $N+2 \rightarrow N+1$ spectrum at the top of Fig. 3b, the spectra are identical. The unmatched state corresponds to an electron tunneling out of a state which is occupied for $N+2$ electrons, but not $N+1$, and therefore it shows occupation of the highest occupied electron orbital state by a single electron. As a function of applied magnetic field, all of the occupied states below this singly-occupied state exhibit Zeeman splitting into two levels, demonstrating that they are doubly degenerate at $B = 0$ as required by Kramers degeneracy in a non-magnetic nanoparticle. We can rule out the presence of any magnetism in the gold nanoparticles also from the fact that we observe that as electrons are added to our nanoparticles by varying gate voltage that they always fill each orbital state with two electrons before occupying the next higher orbital (not shown), and from the absence of any spin blockade effects near zero bias.[28] The absence of magnetism is consistent with the predictions that the exchange interaction in gold is very weak.[25,26] Nevertheless, magnetization measurements of thiol-capped gold nanoparticles have previously suggested the presence of some magnetic character in that system.[29]

As we analyze in more detail below, the magnitudes of the Zeeman splittings in Fig. 3b correspond to effective g factors less than the free-electron value of 2 for each state. In addition, neighboring levels exhibit avoided crossings, rather than simple crossings, as a function of $B$. Both effects are due to spin-orbit coupling[25,30,31,32,33,34,35] and have been observed previously in other metal quantum dots.[3,36,37,38]

(*iii*) The large number of excited states that we can measure in the chemically-formed nanoparticles enables comparisons to the statistical predictions of random matrix theory (RMT) for the distributions of level spacings, for the g factors associated with Zeeman splitting, $g = (\varepsilon_\uparrow - \varepsilon_\downarrow)/(\mu_B B)$, and for the level curvatures as a function of $B$, $k = (\varepsilon_\uparrow + \varepsilon_\downarrow - 2\varepsilon_0)/B^2$. Here $\varepsilon_\uparrow$ and $\varepsilon_\downarrow$ are the Zeeman-split levels of a Kramers doublet, $\varepsilon_0$ is the energy at which they are degenerate at $B = 0$, and $\mu_B$ is the Bohr magneton. Both the g factors and level curvatures are evaluated for $B$ sufficiently large to resolve the Zeeman splitting, but small enough to avoid level crossings where the splitting becomes nonlinear. For convenience in making comparisons to the RMT predictions, we use a nanoparticle whose tunnel couplings to the source and drain electrodes are asymmetric, so that the plot of $dI/dV_{SD}$ vs $V_G$ and $V_{SD}$ (Fig. 4a) contains primarily resonances with just one sign of slope, corresponding to tunneling transitions across just the higher-resistance tunnel junction. In this spectrum only the $N \rightarrow N+1$ resonances ("one-electron excitations") are visible at negative $V_{SD}$, and only the $N+1 \rightarrow N$ transitions ("one-hole excitations") at positive $V_{SD}$, so that in evaluating the level statistics there is no need to sort out overlapping spectra corresponding to different numbers of electrons. We note that sweeping $V_G$ can sometimes cause minor glitches in the evolution of the energy levels due to small changes in the background charge (Fig. 4a), but these should not affect our analyses of level statistics that are performed for fixed $V_G$. Figure 4b shows the magnetic field dependence of the resonances, which indicates that between 0 and 8.6 T the levels shift on average by more than the mean level spacing and typically undergo avoided crossings with neighboring levels. In this sense, 8.6 T is large enough that time-reversal symmetry should be strongly broken.

For a chaotic quantum dot with strong spin-orbit coupling, RMT predicts that the level spacings for $B = 0$ should be described by a Gaussian symplectic ensemble (spin rotation invariance is preserved), with a transition to a Gaussian unitary ensemble for large magnetic fields where time reversal symmetry and spin-rotation symmetry are broken.[22,39] In Figures 5(a,b) we plot the integrated histograms of the energy splittings $\delta$ between neighboring resonances (normalized by the relevant average) for device #2 at $B = 0$ and $B = 8.6$ T. Without any adjustable parameters we find good agreement with the predicted level statistics for the Gaussian symplectic ensemble at $B = 0$ and for the Gaussian unitary ensemble at large field. Quantitatively, we observe a standard deviation equal to 0.33 for the level-spacing distribution of $\delta/\langle\delta\rangle$ at $B = 0$ (GSE predicts $\left[(45\pi/128)-1\right]^{0.5} \approx$ 0.323) and a standard deviation of 0.46 for the distribution $\delta/\langle\delta_{2s}\rangle$ at $B = 8.6$ T (GUE predicts $\left[(3\pi/8)-1\right]^{0.5} \approx 0.422$). Here, $\langle\delta\rangle = 0.23$ meV is the mean level spacing at $B = 0$ and $\langle\delta_{zs}\rangle = 0.12$ meV is the mean spacing of Zeeman-split levels at $B = 8.6$ T.

The g factors for Zeeman splitting near $B = 0$ are listed to the right of Fig. 4b for both the $N \rightarrow N+1$ resonances and $N+1 \rightarrow N$ transitions in device #2. As has been observed previously for noble-metal nanoparticles,[37,38] the g factors fluctuate significantly from level to level within the same nanoparticle, in agreement with expectations for the properties of the highly-oscillatory wavefunctions in a chaotic quantum dot with spin-orbit coupling.[32,33,34,35] In the presence of strong spin-orbit coupling, the g factors are predicted to have contributions from both orbital and spin magnetic moments, in the form[33]

$$\langle g^2\rangle = \alpha\,\frac{l}{L} + \frac{3g_0^2}{2\pi\hbar}\,\tau_{SO}\langle\delta_{zs}\rangle \qquad (1)$$

where $l$ is the mean free path, $L$ is the particle size, $\tau_{SO}$ is the spin-orbit scattering time, $g_0$ is the bulk value of the g factor and $\alpha \approx (6/5)(m^*/m)^{-2}$ for a spherical particle, with $m^* \approx 1.1m$ the bulk effective mass for electrons in gold.[40] For a ballistic particle in the limit of strong spin-orbit scattering, $l \approx L$ and $\tau_{SO}\langle\delta_{zs}\rangle/\hbar \ll 1$, so in this limit it is expected that $\langle g^2\rangle \approx 1$. However, previous measurements from our group[37] and Davidovic and Tinkham[3] on gold particles formed by the clustering of gold atoms during evaporation onto an aluminum oxide surface found much smaller average g factors, ranging from $\langle g\rangle = 0.12$ to $0.45$ (or $\langle g^2\rangle = 0.02$ to $0.20$). For our chemically-synthesized gold nanoparticles, we show in Fig. 5e the g factors measured in 7 different particles, with diameters ranging from 5 to 15 nm, plotted as a function of the local level spacing, defined as the energy difference between the energy level and its nearest neighbor. The g factors range from nearly zero to nearly 2, with no apparent dependence on the level spacing. The overall averages and standard deviations for these samples are $\langle g\rangle = 0.85$, $\Delta g = 0.41$ and $\langle g^2\rangle = 0.88$, $\Delta g^2 = 0.79$. These magnitudes of the g factors are therefore in much better agreement with the prediction of Eq. (1) for ballistic particles with strong spin-orbit coupling than the previous measurements on evaporated nanoparticles. In Fig. 5f, we plot the integrated histogram of these measured g factors and in Fig. 5c we plot the same quantity for the g factors just from device #2, with comparisons given to the appropriate RMT prediction[32,33] with $\langle g^2\rangle$ determined by squaring and averaging the measured values of $g$ (i.e., there are no free fitting parameters). Fig. 5d shows the measured distribution of the level curvature, $k$, with a comparison to the RMT prediction for this quantity, with no adjustable parameters once $\langle|k|\rangle$ is evaluated.[35,41] In all cases, the RMT predictions for the forms of the distributions are excellent. (We note, however, that the predicted distributions of level curvatures for other RMT ensembles[41] are sufficiently similar that our data do not distinguish between GSE statistics and the other ensembles for this quantity.) RMT also predicts that in the limit of strong spin-orbit scattering,[35]

$$\langle|k|\rangle = \frac{2\sqrt{2}\mu_B^2}{9\langle\delta_{zs}\rangle}\langle g^2\rangle. \qquad (2)$$

For device #2, with $\langle g^2 \rangle = 0.97$ and $\langle \delta_{zs} \rangle = 0.12$ meV, this predicts $\langle |k| \rangle = 8.6$ $\mu$eV/T$^2$, in reasonable agreement with our measured value $\langle |k| \rangle = 7.8$ $\mu$eV/T$^2$.

The difference between the previously-measured magnitudes of the g factors in evaporated Au nanoparticles[3,37] and our current results on chemically-formed Au nanoparticles could result from different degrees of disorder. If $l << L$ in the evaporated nanoparticles, this would decrease the magnitude of the orbital-moment term (the first term on the right in Eq. (1)) and predict smaller average g factors. However, very small mean free paths ($\sim 0.3$ nm) would seem to be required to reconcile the experimental results for the evaporated particles with the theory. Another difference between these experiments occurs at the surface of the nanoparticles; the evaporated nanoparticles were covered by aluminum oxide tunnel junctions rather than an organic monolayer. It may be worth considering whether hybridization with surface states or strong scattering at the gold-oxide interface could suppress the orbital contribution to the g factor in the evaporated nanoparticles.

In summary, we have demonstrated that individual chemically-synthesized metal nanoparticles can be incorporated by self-assembly into a single-electron transistor by means of a functionalized organic monolayer. The devices have excellent mechanical and electrical stability, sufficient to allow detailed spectroscopy of the quantum-mechanical electron-in-a-box energy levels in the nanoparticle at dilution-refrigerator temperatures. For nearly-spherical gold nanoparticles 5-15 nm in diameter, we find that the properties of the quantized states are strongly affected by spin-orbit coupling, in good accord with the predictions of random matrix theory for a chaotic quantum dot. Also, the energy-level spectra are not "scrambled" by the addition of electrons, indicating that in these particles the variation in the strength of electron-electron interactions between states are negligible, so that a constant interaction model is an adequate approximation. Recent advances in techniques for chemical synthesis have enabled nanoparticles and nanoparticle heterostructures to be made from a wide variety of materials, with excellent control over particle size, shape and composition.[42,43,44,45] The use of electron tunneling spectroscopy to measure the energy-level spectra in chemically-formed nanoparticles can therefore provide a powerful means for exploring the quantum mechanics of electrons and their interactions in many materials and new nanostructure geometries.

We thank Piet Brouwer and Eduardo Mucciolo for helpful discussions and Mick Thomas for technical assistance. We acknowledge funding from the NSF (DMR-0605742, CHE-0403806, and through use of the Cornell NanoScale Facility/NNIN).

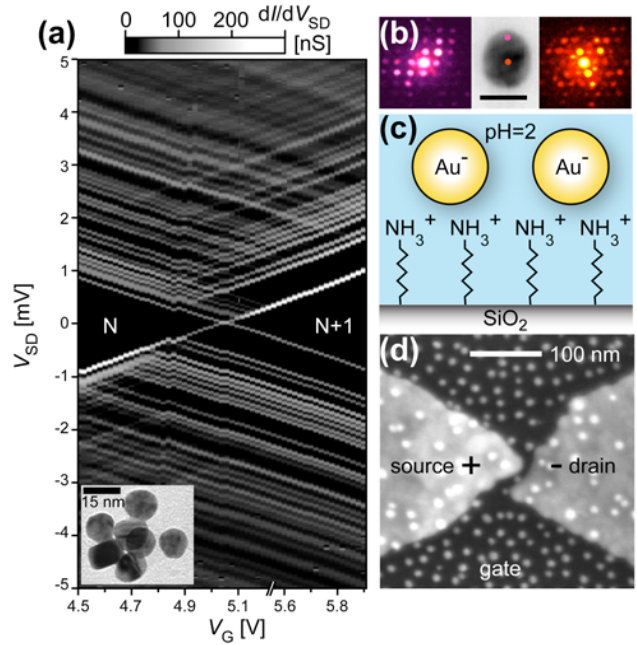

**Figure 1**. (a) Discrete electronic level spectrum of a gold nanoparticle at $B = 1.5$ T at dilution-refrigerator temperatures, $T \leq 90$ mK (device #1 from nominally 15-nm-diameter colloid). Inset: Transmission electron microscope image of chemically synthesized gold nanoparticles with mean diameter 12 nm. (b) Collimated-electron-beam diffraction images obtained from two different spots within a single 10-nm-diameter nanoparticle suggest that the nanoparticles are polycrystalline (scale bar is 10 nm). (c) During our self-assembly process, negatively charged nanoparticles in acidic aqueous suspension are attracted to partially-protonated amino groups in an organic monolayer. (d) Scanning electron microscope image of a finished device in which only one nanoparticle contributes to current flow between source and drain electrode (device #2, from nominally 10-nm-diameter gold colloid).

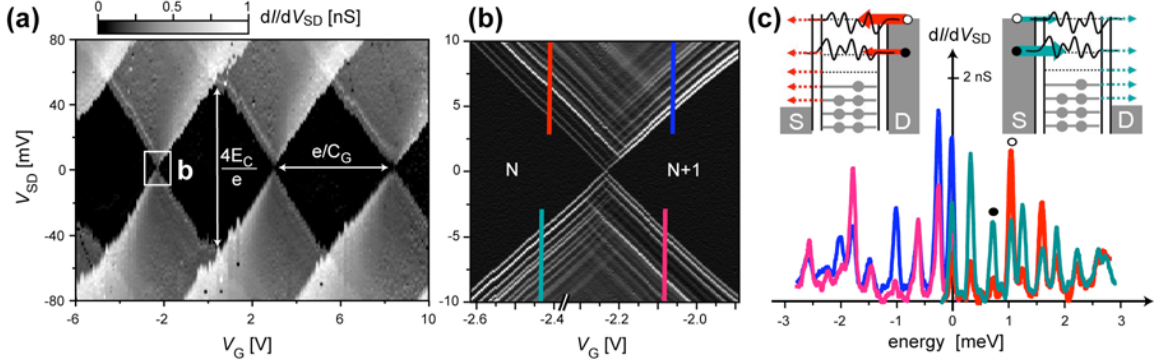

**Figure 2.** (a) Differential conductance as a function of $V_{SD}$ and $V_G$ over large voltage ranges for a gold-nanoparticle single-electron transistor at $B = 0$ and electron temperature $\leq 90$ mK (device #3, from 9-nm-diameter gold colloid). The charging energy $E_C$ and gate capacitance $C_G$ are extracted from the size of the regions in which the conductance is zero due to Coulomb blockade. (b) Differential conductance obtained near one degeneracy point for this sample. (c) Line cuts from b) with $V_{SD}$ converted to energy relative to the ground state. Each resonance appearing at positive bias also appears at negative bias, but with varying conductance amplitudes. Insets: Schematic illustration of two quasi-random wavefunctions within a particle. The lower energy state can have a stronger tunnel coupling to the source electrode while the higher state can have stronger coupling to the drain, so that their conductance ratio changes if the bias is reversed.

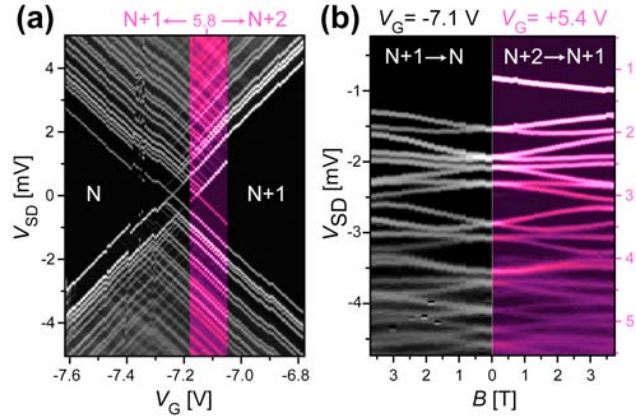

**Figure 3**. (a) Differential conductance obtained from $N \rightarrow N+1$ and $N+1 \rightarrow N$ electron-number tunneling transitions (grayscale), overlaid with the conductance obtained from $N+1 \rightarrow N+2$ and $N+2 \rightarrow N+1$ transitions in the same device (pink) (device #1). Both spectra are for $B = 0$ and electron temperature $\leq 90$ mK. (b) Magnetic field dependence of these two spectra for constant $V_G$, obtained near the two degeneracy points in panel (a). The ground-state to ground-state transition for $N+2 \rightarrow N+1$, at the top right, shows only one state of a Kramer's doublet, demonstrating that $N+2$ is odd.[1]

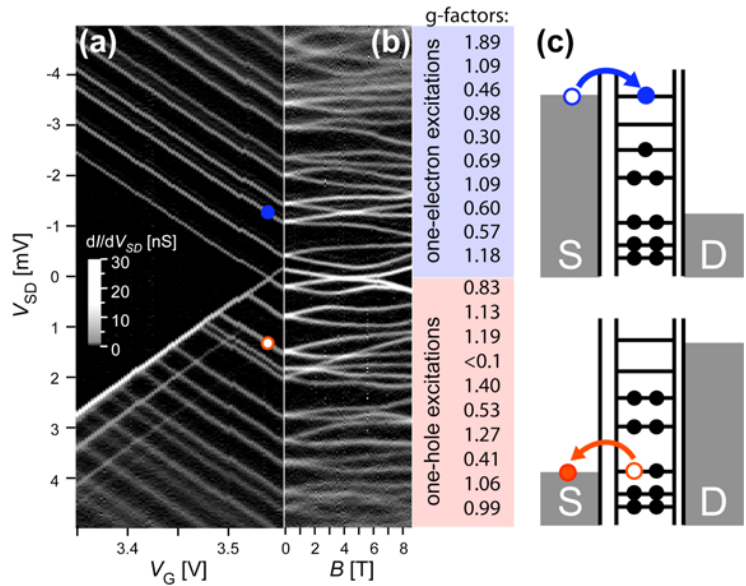

**Figure 4.** (a) Differential conductance at $B = 0$ T for device #2 (from nominally 10-nm-diameter gold colloid), which has asymmetric tunnel barriers. A resonance for which a one-electron excitation (one-hole excitation) is the rate-limiting step is marked by the blue (red) symbol. (b) Magnetic field dependence at $V_G = 3.552$ V. The g factors for Zeeman splitting near $B = 0$ are listed to the right of each level. (c) Energy-level diagrams corresponding to the tunneling transitions labeled by the red and blue symbols in (a).

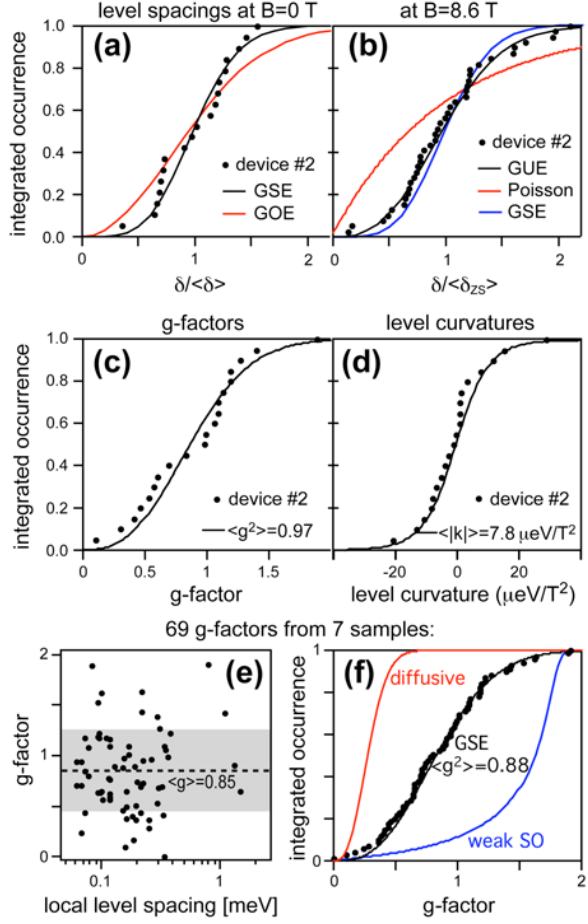

**Figure 5**. (a) Symbols: Integrated probability distribution of level spacings $\delta$ at $B = 0$ T for device #2, taken from Fig. 4b and normalized by the mean level spacing $\langle \delta \rangle = 0.23$ meV. Comparisons are made to the level-spacing statistics predicted by the Gaussian symplectic random matrix ensemble (GSE) and Gaussian orthogonal ensemble (GOE). (b) Integrated probability distribution of level spacings at $B = 8.6$ T for device #2 taken from Fig. 4b and normalized by the mean spacing of Zeeman-split levels $\langle \delta_{zs} \rangle = 0.12$ meV. Comparisons are made to the Gaussian unitary random matrix ensemble (GUE), the GSE, and a random Poisson distribution. (c) Integrated probability distribution for the g factors from device #2, compared to the prediction of RMT for strong spin-orbit coupling with $\langle g^2 \rangle = 0.97$. (d) Integrated probability distribution for the measured level curvatures $k$ for device #2, with comparison to the prediction of RMT for strong spin-orbit coupling with $\langle |k| \rangle = 7.8$ $\mu$eV/T². (e) 69 g factors measured from 7 devices, plotted as a function of the local level spacing (see text). The mean and standard deviation are indicated by the dashed line and gray background, respectively. (f) Integrated probability distribution of the g factors from these seven devices, with comparisons to the distributions predicted by RMT[32,33] in the limit of strong spin orbit coupling for $\langle g^2 \rangle = 0.88$ (consistent with ballistic transport) and $\langle g^2 \rangle = 0.1$ (consistent with diffusive transport), as well as a typical distribution for weaker spin-orbit coupling corresponding to $\langle g \rangle = 1.5$.